\documentclass[12pt,titlepage]{utarticle}

\DeclareMathSymbol{\square}{\mathord}{AMSa}{"03}
\DeclareMathSymbol{\rightsquigarrow}{\mathrel}{AMSa}{"20}

\newdimen\tableauside\tableauside=1.0ex
\newdimen\tableaurule\tableaurule=0.4pt
\newdimen\tableaustep
\def\phantomhrule#1{\hbox{\vbox to0pt{\hrule height\tableaurule width#1\vss}}}
\def\phantomvrule#1{\vbox{\hbox to0pt{\vrule width\tableaurule height#1\hss}}}
\def\sqr{\vbox{%
  \phantomhrule\tableaustep
  \hbox{\phantomvrule\tableaustep\kern\tableaustep\phantomvrule\tableaustep}%
  \hbox{\vbox{\phantomhrule\tableauside}\kern-\tableaurule}}}
\def\squares#1{\hbox{\count0=#1\noindent\loop\sqr
  \advance\count0 by-1 \ifnum\count0>0\repeat}}
\def\tableau#1{\vcenter{\offinterlineskip
  \tableaustep=\tableauside\advance\tableaustep by-\tableaurule
  \kern\normallineskip\hbox
    {\kern\normallineskip\vbox
      {\gettableau#1 0 }%
     \kern\normallineskip\kern\tableaurule}%
  \kern\normallineskip\kern\tableaurule}}
\def\gettableau#1 {\ifnum#1=0\let\next=\null\else
  \squares{#1}\let\next=\gettableau\fi\next}

\begin{document}

\preprint{
 HUB-EP-97/13\\
 {\tt hep-th/9702179}\\
}

\title{More on $N=1$ Self-Dualities and Exceptional Gauge Groups}
\author{Andreas Karch
 \thanks{Work supported by DFG}
 \oneaddress{
  \\
  Humboldt-Universit\"at zu Berlin\\
  Institut f\"ur Physik
  D-10115 Berlin, Germany\\
  {~}\\
  \email{karch@qft1.physik.hu-berlin.de}
 }
}
\date{February 25, 1997}

\Abstract{
Starting from a generalization of a recent result on self-duality \cite{boston}
we systematically analyze self-dual models. We find a criterion to judge
whether a given model is self-dual or not. With this tool we construct some
new self-dual pairs, focussing on examples with exceptional gauge groups. 
}

\maketitle

\section{Introduction}

In the last few years it has become clear that many $N=1$ supersymmetric
gauge theories in four dimensions exhibit the phenomenon of duality 
(\cite{seiberg}, \cite{classical}-\cite{boston}).
Within these huge amount of known examples there
is a class of models that deserves special attention: the self-dual
ones. Self-duality in this context means that the magnetic gauge
group is the same as the electric one, but they do not describe the
same UV physics, since the magnetic theory has additional gauge
singlets and superpotential terms.

These self-dual models have some very interesting features. One
on hand they look very simple. The 't Hooft anomaly matching conditions
are always satisfied in a very trivial fashion.
They are very helpful in building duals with matter
in a tensor representation of the gauge group
and
without an electric tree level superpotential \cite{sp} and
they are almost the
only information we got so far 
about exceptional gauge groups other than $G_2$ (\cite{ramond, myfirst}).
On the other hand they were shown to be related to
the existence of exactly marginal operators \cite{leigh} and
this might make them somewhat easier to study.

Recently several new self-dualities have been found in \cite{boston}
for $SU$ groups with various higher rank antisymmetric tensors
and for some $Spin$ groups with spinors. In this paper we
will generalize their $Spin$ self-dualities, find a tool
to construct new self-dual models and finally use it
to get some new information about exceptional gauge groups.   

Section 2 contains a review of the $Spin$ models of \cite{boston}
and a series of similar dualities. We will find the first
indications that there should be a criterion for possible
self-duality involving the object $\Delta = \mu_{gauge} - \mu_{matter}$
like the one found in \cite{confine} for the construction of s-confining
theories.

In Section 3 we will construct this tool and show why it
works and what its limitations are. In Section 4 we then use this
new tool to get information about exceptional gauge groups
and to construct other new self-dual models. 

We will conclude in section 5.

\section{The $Spin$ series of \protect \cite{boston}}

In \cite{boston} the authors discovered several new selfdual models. Within those was a series of $Spin(N)$ models with $N-4$ vectors. The important information about those models is contained in the following table, taken from \cite{boston}.

\begin{displaymath}
\begin{array}{c|c|l}
{\rm group} & {\rm content}     &       {\rm ``mesons"} \\ \hline \hline
Spin(4) & (8,8,0)  & \sim SU(2)\times SU(2) \; \mbox{with}\;
8(\tableau{1},1) + 8 (1,\tableau{1}) \\
Spin(5) & (8,1) & \sim Sp(4) \; \mbox{with}\;  8\, \tableau{1} + \tableau{1 1} \\
Spin(6) & (4,4,2)  & \sim SU(4) \; \mbox{with}\;
4(\tableau{1} +\overline{\tableau{1}}) + 2\, \tableau{1 1} \\
Spin(7)   & (4,3)         &       s^2,s^2q,s^2q^2,s^2q^3  \\
Spin(8)   & (4,0,4)       &       s^2,s^2q^2,s^2q^4 \\
Spin(8)   & (2,2,4)  &  s^2,s^2q^2,s^2q^4,c^2,c^2q^2,c^2q^4,scq,scq^3 \\
Spin(8)   & (3,1,4) & s^2, s^2q^2,s^2q^4,c^2,c^2q^4,scq,scq^3 \\
Spin(9)   & (2,5) &               s^2,s^2q,s^2q^2,s^2q^3,s^2q^4,s^2q^5 \\
Spin(10)  & (2,0,6)       &       s^2q,s^2q^3,s^2q^5 \\
Spin(10)  & (1,1,6)  & s^2q,s^2q^5,c^2q,c^2q^5,sc,scq^2,scq^4,scq^6 \\
Spin(11)  & (1,7) &               s^2q,s^2q^2,s^2q^5,s^2q^6 \\
Spin(12)  & (1,0,8)       &       s^2q^2,s^2q^6 \\
\end{array}
\end{displaymath}
\begin{center}
{\bf Table 1:} Selfdual models of \cite{boston}
\end{center}

The first column denotes the electric gauge group, the second column denotes the matter content,
(number of spinors, number of vectors) for odd $N$ and (number of spinors, number of conjugate spinors, number of vectors) for even $N$. Since by definition of what we mean by self-duality, the magnetic gauge group is the same as the electric
one, to fully specify the model, it is enough to list the operators becoming 
fundamental gauge singlets on the magnetic side and whose corresponding
magnetic operator gets removed via the superpotential. This information is contained
in the third column.
\footnote{Here and in the following we denote magnetic fields by small letters
and electric fields by capital letters. We use $q,Q$ as a name 
for fields in the fundamental representation 
of the gauge group (the vector in the $Spin$ cases), $s,S$ for spinors and
$c,C$ for conjugate spinors.}

The superpotential is constructed in the obvious way.

For example, the last line contains the data of the following dual pair:

electric:

\begin{center}

\begin{tabular}{|c||c||c|c|c|}
\hline
&$Spin(12)$&$SU(8)$&$U(1)$&$U(1)_R$\\
\hline
\hline
$Q $&$12$&8&$1$&$\frac{1}{6}$\\
\hline
$S $&$32$&1&$-2$&$\frac{1}{6}$\\
\hline
\end{tabular}

\end{center}

magnetic:

\begin{center}

\begin{tabular}{|c||c||c|c|c|}
\hline
&$Spin(12)$&$SU(8)$&$U(1)$&$U(1)_R$\\
\hline
\hline
$q $&$12$&$8$&$1$&$\frac{1}{6}$\\
\hline
$s $&$32$&1&$-2$&$\frac{1}{6}$\\
\hline
$M_4$&$1$&$\tableau{1 1}$&$-2$&$\frac{2}{3}$\\
\hline
$M_8$&$1$&$\overline{\tableau{1 1}}$&$2$&$\frac{4}{3}$\\
\hline
\end{tabular}

with: $W= M_4 s^2 q^6 + M_8 s^2 q^2$.
\end{center}
 
The gauge invariant operator $S^2 Q^6$ gets mapped to $M_8$,
$S^2 Q^2$ to $M_4$. $s^2 q^6$ and $s^2 q^2$ are removed via
the equations of motion for $M_4$ and $M_8$. All other
independent gauge invariant operators are the same on both
sides and get mapped into each other.

As indicated, the first three entries in Table 1 are equivalent
to other theories which were allready known to be self-dual.
This is another check for the consistency of the models.
Giving a vev to one of the vectors, one can flow from a $Spin(N)$ dual in the
list to a $Spin(N-1)$ one. Giving a mass to one of the vectors, one flows
to a s-confining model (\cite{confine}). All the s-confining $Spin$ theories
have been classified by the same authors in \cite{confine}. The fact, that
the list of those s-confining theories contains another similar series of
s-confining theories hints that there is also another series of self-dual
$Spin$ models. In fact it is easy to check, that the following models are also
self-dual:

\begin{displaymath}
\begin{array}{c|c|l}
{\rm group} & {\rm content}     &       {\rm ``mesons"} \\ \hline \hline
Spin(8)   & (4,4,0)         &       s^2,s^2c^2,s^2c^4  \\
Spin(9)   & (4,1)       &        \\
Spin(10)   & (2,2,2)  &  sc^3q,s^3cq\\
Spin(11)   & (2,3) &               s^2q,s^2q^2,s^6q,s^6q^2 \\
Spin(12)  & (2,0,4)       &       s^2q^2,s^6q^2 \\
Spin(12)  & (1,1,4)  & s^2q^2,c^2q^2,s^4c^2q^2,s^2c^4q^2 \\
Spin(13)  & (1,5) &               s^2q^2,s^2q^3,s^6q^2,s^6q^3 \\
Spin(14)  & (1,0,6)       &       s^2q^3,s^6q^3 \\
\end{array}
\end{displaymath}

As evidence for those dualities we take as usual, that the 't Hooft matching
conditions for the global symmetries are satisfied and that we
can consistently perturb the dual pairs along flat directions
to another consistent dual pair.
We only checked
the vector flat directions, 
which connect the theories listed in the table with each other. But there is no indication that the spinors should make trouble.
The final $Spin(8)$ theory with $(4,4,0)$ is equivalent to the $Spin(8)$ theory
with $(4,0,4)$ of \cite{boston} by triality.

Mass terms again perturb to s-confining theories, and these two series together now reproduce allmost all the
known s-confining models. This close connection between s-confining
models and self-dual models is kind of surprising. In the next section
we will argue that this is due to the fact that both phenomena
can be analyzed by looking at the expression $\mu_{matter}-\mu_{gauge}$.
Here $\mu$ stands for the quadratic index of the gauge group. 
$\mu_{matter}$ is the sum of all the indices of all the matter fields in the various representations and $\mu_{gauge}$ is the index of the gauge fields (the
index of the adjoint). We normalize $\mu$ to be 1 for a fundamental $SU$ 
representation.

\section{A Tool for constructing self-dual models}

The quantity $\Delta \equiv \mu_{matter}-\mu_{gauge}$ has already proved
to be useful in several cases. \cite{confine} showed that $\Delta=2$
indicates a s-confining theory. Even though this alone was neither
necessary nor sufficient, it proved to be very useful for their work. 
They were able to classify all the s-confining models with little more
work than ckecking the above condition: together with their
second condition obtained by studying different regions on the moduli space the index constraint became necessary. This way
they created a list of all possible candidate theories and had
to check them by their second criterion (going along flat directions).
Similary, $\Delta=0$ indicates a quantum smoothed out moduli space
and $\Delta=-2$ a non-perturbative superpotential created by a single
instanton effect.

We would like to argue that in a similar way $\Delta=k$ and $k$ a divisor
of $\mu_{gauge}$ is an indication for self-duality. This is 
again neither a necessary nor a sufficient condition. Nevertheless
it proved to be a useful tool, since it allowed us to
construct several new self-dual models, which we will present
in the next section.

Before we explain why the above condition is important let us make
some comments. Of all the self-dual models without a tree level superpotential
 we know of (\cite{seiberg,leigh,ramond,sp,myfirst,boston} )
only the odd $Spin(N)$ models from above, the $Sp(2n)$ theories with odd n
of \cite{sp} and some of
the models from the next section do not meet the requirement, that
$\Delta$ is a divisor of $\mu_{gauge}$. And all those models can
be obtained by giving a vev to some field in a theory that does meet
the requirement \footnote{To see this in the Sp(2n) theory
of \cite{sp} consider giving a vev to the asym. tensor A of the form
$<A> = i \sigma_2 v \otimes \left (
	   \begin{array}{cccc}
			 1 & & & \\
				& 0 & & \\
				       & & \ddots & \\
					    & & & 0 
	  \end{array} \right)$.
This breaks the gauge group to $Sp(2(n-1)) \times SU(2)$. Since the operators
involving only A are mapped onto themselves on the magnetic side, exactly
the same happens to the magnetic theory. The $SU(2)$ decouples and one
obtains the self-dual $n-1$ theory from the $n$ theory.
}. Probably there are other self-dual models
not related at all to our requirement. But as will become clear
soon, their structure is much more complicated and so no one has
guessed them so far. We will have to wait for a better understanding
of the mechanisms of duality in $N=1$ theories before we can try
to systematically analyze them.

The case $k=2$ deserves special attention. There are many
theories meeting this requirement.
Luckily the $k=2$ case is exactly the case discussed in \cite{confine}
that also indicates s-confinement. So most of the work in this case has
been done. All the $Spin(N)$ theories from the last section
have $\Delta=4$. For even $N$ this is a divisor of $\mu_{gauge}=4N-4$.
And the odd ones can be obtained from the even ones by higgsing a
vector. Removing a vector by giving it a mass reduces $\Delta$ to
2 and gives an s-confining model.

Now let us explain why the requirement $\Delta=k$ and $k$ a divisor of
$\mu_{gauge}$ is important for self-duality. If one assigns the
same r-charge $r$ to all the matter fields, for an anomaly free r-symmetry
this has to be $r=1-\frac{\mu_{gauge}}{\mu_{matter}}$. The above
requirement then translates to the requirement $r=\frac{1}{n}$ for
some integer $n$ ($n=\frac{\mu_{gauge}}{k}+1$). This is important for
self-duality for two reasons.

The first is a very technical point of view. To satisfy the 't Hooft matching
conditions for the global r-symmetry with the magnetic and the electric gauge group being the same
is almost impossible unless the additional gauge singlets
on the magnetic side have either r-charge 1 (which means that they do not
contribute at all) or always come in pairs of r-charge $1+a$ and $1-a$
(in which case their contribution cancels). If the r-charge
of all fields is just $\frac{1}{n}$ such operators are very easy to construct:
a composite built out of n fields has r-charge 1 and composites built out of
$n+m$ fields and $n-m$ fields (m integer and less than n) form an $(1+a,1-a)$ pair. Even though it is in principal not impossible to 
construct such operators in a different way or to satisfy the
r-anomaly with a different mechanism, this is the easiest way to
do it and explains, why all the models found so far are either
constructed this way or derived from a model constructed this way.

It should allready be clear at this point, why we have trouble
with the theories with tree level superpotential: in 
the presence of a tree level superpotential term on the electric 
side we loose the freedom to freely choose the r-charge of the
fields to be equal. We have to fix it to whatever is preserved
by the superpotential.

The second argument uses the deep connection between self-dual models and
exactly marginal operators. Following the lines of \cite{leigh} it
is natural to conjecture
that
in every self-dual model there is an exactly marginal operator
that parametrizes a fixed line smoothly connecting the electric and
the magnetic theory. The way this marginal operator arises is the
following: Consider the magentic theory and add a mass term for
the mesons (that is what in the case for self-dual SUSY QCD was
called Supersymmetric Quantum Chromesodynamics in \cite{leigh})
The $m=0$ limit clearly reproduces the magnetic theory, 
$m \rightarrow \infty$ gives back the electric theory. Integrating
out the meson at intermediate values of $m$ generates
an operator in the superpotential which is exactly
marginal (if there are different types of mesons, a linear combination
of the operators created this way is conjectured to be marginal). This structure arose in their case from the close
connection to $N=2$, but the conjecture is supposed to be true
even for models with no $N=2$ origin. Hence we can find 
self-dual models by searching for theories with exactly marginal operators.
We will show that with the r-charge assignment and
meson content introduced above
(which was only possible if the $\Delta$ constrained was satisfied)
the model obtained this way has indeed a fixed
line instead of just a fixed point and hence an exactly marginal operator.

To establish this, it is necessary to study the $\beta$-functions of the
gauge coupling and the coupling of a linear combination of the operators
associated with the various
mesons. These operators are of the form: ``gauge invariant operator that
becomes a elementary magnetic gauge singlet'' times ``the gauge invariant
operator it removes on the magnetic side via its equation of motion''.
For example in SUSY QCD we are looking for $(Q \tilde{Q})^2$ (see \cite{leigh})
and in the $Spin(12)$ theory of \cite{boston} discussed in section 2 for
$(S^2 V^2) (S^2 V^6)$, all the indices being contracted.

Since these operators appear in the superpotential they have to
have r-charge 2 and hence also any linear combination of them.
They all have $2n$ fields (since each field has r-charge $\frac{1}{n}$),
and invariance of the superpotential under the various anomaly
free $U(1)$s tells us that the number $\#_i$ of fields of type $i$ 
in anyone of the products in the linear combination that makes up 
our operator is $\#_i=\frac{2 \mu_i }{\Delta}$.

Existence of a fixed line is equivalent to the fact that the scaling
coefficents of the gauge coupling $A_g$ and for the superpotential 
coupling of the operator $A_{h}$
are linearly dependend. They are
\begin{eqnarray*}
A_g&=& - 3 \mu_{gauge}+\mu_{matter} - \sum_i \mu_i \gamma_i \\
   &=& \mu_{gauge} \cdot \frac{3-2n}{n-1} - \sum_i \mu_i \gamma_i\\
A_h&=& 2n-3 + \frac{1}{2} \sum_i \#_i \gamma_i \\
   &=& -(3-2n)+\sum_i \mu_i \frac{n-1}{\mu_{gauge}} \gamma_i\\
\end{eqnarray*}
where $\gamma_i$ denotes the anomalous mass dimension of field type $i$ and
we used $\Delta=\frac{\mu_{gauge}}{n-1}$ various times. These
are now obviously linearly dependend, since $A_g = - \frac{\mu_{gauge}}{n-1}
A_h$.

\section{New Self-Dual Models}
\subsection{Exceptional Gauge Groups}

One of the big unsolved problems in $N=1$ dualities is finding
a dual description for models with exceptional gauge groups. All the
progress made so far is based on self-dual models (\cite{ramond}, \cite{myfirst}). With the tool described in the last section, we were able to construct some
new exceptional self-dualities. In the short notation introduced
in section 2 they are:

\begin{displaymath}
\begin{array}{c|c|l}
{\rm group} & {\rm content}     &       {\rm ``mesons"} \\ \hline \hline
E_6  & (3,3)  & q \tilde{q}, q_a^2 \tilde{q}_a^2 \\
F_4  & (4) &    q_s^2, q^6 (\tableau{2 2 2})            \\
G_2  & (6)       &  q_a^3      \\
\end{array}
\end{displaymath}

With the following $\Delta$ values
\begin{eqnarray*}
\Delta(E_6)&=&6 \mu_{27} - \mu_{78}=6 \cdot 6 -24 =\frac{1}{2} \mu_{78}\\
\Delta(F_4)&=&4 \mu_{26} - \mu_{52}=4 \cdot 6 - 18 =\frac{1}{3} \mu_{52}\\
\Delta(G_2)&=&6 \mu_{7} - \mu_{14}=6 \cdot 2 - 8 =\frac{1}{2} \mu_{14}\\
\end{eqnarray*}
the three clearly qualify for self-duality by the arguments of the
previous section. There is a one to one mapping of gauge invariant
operators and the 't Hooft anomaly matchings are satisfied.
As further evidence we checked some of the flat directions:

Consider perturbing the $F_4$ model along the flat direction described by 
$$M^{111}= d^{\alpha \beta \gamma} Q^1_{\alpha} Q^2_{\beta} Q^3_{\gamma},$$
$\alpha, \beta, \gamma$ being color indices, 1 a flavor index and
d the 3-index totally symmetric invariant tensor of $F_4$. That is,
we give one flavor a vev in such a way, that only the invariant built
with $d$ gets a non-zero vev. This higgses the electric gauge
theory to $Spin(8)$ with $(3,3,3)$. Since the invariant is
mapped to the magnetic one built in exactly the same way, the
magnetic gauge group is also $Spin(8)$ with $(3,3,3)$. This
is indeed again a new consistent self-dual pair with
\begin{eqnarray*}
\Delta(Spin(8))&=&9 \mu_{8} - \mu_{28}=18-12 =\frac{1}{2} \mu_{28}.\\
\end{eqnarray*}
and the following content:

\begin{center}

\begin{tabular}{|c||c||c|c|c|c|c|c|}
\hline
&$Spin(8)$&$SU(3)_V$&$SU(3)_S$&$SU(3)_C$&$U(1)$&$U(1)$&$U(1)_R$\\
\hline
\hline
$Q$&$8_V$&$3$&$1$&$1$&$1$&$1$&$\frac{1}{3}$\\
\hline
$S$&$8_S$&$1$&$3$&$1$&$-1$&$0$&$\frac{1}{3}$\\
\hline
$C$&$8_C$&$1$&$1$&$3$&$0$&$-1$&$\frac{1}{3}$\\
\hline
\end{tabular}

$\updownarrow$

\begin{tabular}{|c||c||c|c|c|c|}
\hline
&$Spin(8)$&$SU(3)_D$&$U(1)$&$U(1)$&$U(1)_R$\\
\hline
\hline
$q$&$8_V$&$3$&$1$&$1$&$\frac{1}{3}$\\
\hline
$s$&$8_S$&$3$&$-1$&$0$&$\frac{1}{3}$\\
\hline
$c$&$8_C$&$3$&$0$&$-1$&$\frac{1}{3}$\\
\hline
$M_2$&$1$&&$2$&$2$&$\frac{2}{3}$\\
$N_2$&$1$&$\tableau{2}$&$-2$&$0$&$\frac{2}{3}$\\
$P_2$&$1$&&$0$&$-2$&$\frac{2}{3}$\\
\hline
$M_4$&$1$&&$-2$&$-2$&$\frac{4}{3}$\\
$N_4$&$1$&$\tableau{2 2}$&$+2$&$0$&$\frac{4}{3}$\\
$P_4$&$1$&&$0$&$+2$&$\frac{4}{3}$\\
\hline
\end{tabular}

with: $W= M_2 c_a^2 s_a^2 + N_2 q_a^2 c_a^2 + P_2 q_a^2 s_a^2 +
M_4 q_s^2+ N_4 s_s^2 +P_4 c_s^2$ .
\end{center}

On the magnetic side only the diagonal subgroup of the electric 
$SU(3) \times SU(3) \times SU(3)$ global flavor rotations is
visible on the fundamental magnetic fields. The full
symmetry only appears as an accidental symmetry in the far IR. This
is by now a well known phenomenon, that appeared over and over again
(\cite{acci,myfirst,boston}). It is not surprising to
see it pop up here: the diagonal $SU(3)$ is the one
inherited from the $F_4$ theory. On the electric side the only objects
charged under this $SU(3)$ are the 27s, which split up into three
objects under the broken gauge group, which can be rotated independently.
On the magnetic side there are also the mesons transforming under the
global flavor rotations. They are not affected by the higgsing and
there is no reason why they should transform under three seperate rotations.
Their superpotential contribution prevents one from rotating the three
different 8s independently. Only in the IR, where it only makes sense to
talk about operators, the additional rotations show up, transforming
mesons and composite gauge invariants into each other.
Of the invariants like $V^2 S^2$ only the part transforming like
$\tableau{2 2}$ under the diagonal
$SU(3)$ is mapped to a magnetic gauge singlet, the
other components are composite gauge invariants.
 
Similarly one can perturb the $G_2$ theory along its $SU(3)$ flat direction.
We get a self-duality for $SU(3)$ with 5 flavors. For the same
reasons as in the $F_4$ case we expect to see only the
diagonal subgroup of the global $SU(5) \times SU(5)$ flavor rotations.
This leads us to the following self-dual pair:

\begin{center}

\begin{tabular}{|c||c||c|c|c|c|}
\hline
&$SU(3)$&$SU(3)_L$&$SU(3)_R$&$U(1)$&$U(1)_R$\\
\hline
\hline
$Q$&$3$&$5$&$1$&$1$&$\frac{2}{5}$\\
\hline
$\tilde{Q}$&$\overline{3}$&$1$&$5$&$-1$&$\frac{2}{5}$\\
\hline
\end{tabular}

$\updownarrow$

\begin{tabular}{|c||c||c|c|}
\hline
&$SU(3)$&$SU(5)_D$&$U(1)_R$\\
\hline
\hline
$q$&$3$&$5$&$\frac{1}{3}$\\
\hline
$\tilde{q}$&$\overline{3}$&$5$&$\frac{1}{3}$\\
\hline
$M$&$1$&$\tableau{1 1}$&$\frac{2}{3}$\\
\hline
$B$&$1$&$\overline{\tableau{1 1}}$&$\frac{4}{3}$\\
\hline
\end{tabular}

with: $W= M (q^3 + \tilde{q}^3) + B q \tilde{q}$
\end{center}

This new self-duality can be generalized to SUSY QCD with arbitrary gauge
group and $N_c+2$ flavors. The models with even $N_c$ satisfy our
$\Delta$ condition ($\Delta=2N_c+4-2N_c=4$, $\mu_{gauge}=2N_c$)
and the odd values can be obtained by higgsing. Only
the diagonal subgroup of flavor rotations is visible in the magnetic theory.
The antisymmetric part of the meson and one baryon appear as elementary
gauge singlet fields on the magnetic side, the symmetric part of the
meson and the other baryon are composite fields.

There is another piece of evidence for the $G_2$ model:
In addition to the $Spin$ models of the two series in section 2, there
is another self-dual model for $Spin(7)$:

electric:

\begin{center}

\begin{tabular}{|c||c||c|c|c|}
\hline
&$Spin(7)$&$SU(6)$&$U(1)$&$U(1)_R$\\
\hline
\hline
$Q $&$7$&1&$6$&$\frac{2}{7}$\\
\hline
$S $&$8$&6&$-1$&$\frac{2}{7}$\\
\hline
\end{tabular}

\end{center}

magnetic:

\begin{center}
\begin{tabular}{|c||c||c|c|c|}
\hline
&$Spin(7)$&$SU(6)$&$U(1)$&$U(1)_R$\\
\hline
\hline
$q $&$7$&$1$&$6$&$\frac{2}{7}$\\
\hline
$s $&$8$&6&$-2$&$\frac{2}{7}$\\
\hline
$M_3$&$1$&$\tableau{1 1}$&$+4$&$\frac{6}{7}$\\
\hline
$M_4$&$1$&$\overline{\tableau{1 1}}$&$-4$&$\frac{6}{7}$\\
\hline
\end{tabular}

with: $W= M_4 s^2 q + M_3 s^4$.
\end{center}
mapping $S^4$ to $M_4$ and $S^2 Q$ to $M_3$.

This theory also s-confines after giving a mass to one vector. It is
not a trivial consequence of the $Spin(8)$ models and triality, since
triality is hidden in the magnetic $Spin(8)$ theories (it appears only
in the infrared). After following this theory along
a spinor flat direction we reproduce the $G_2$ theory. Following
the vector flat direction we get $Spin(6)=SU(4)$ with 6 flavors and
hence we again hit SUSY QCD with $N_c+2$ flavors,
which we allready showed to be self-dual.

\subsection{Self-dual models with symmetric tensors}

Even though self-duality created a lot of knowledge about $SU$ theories
with antisymmetric tensor representations, much less has been achieved
in theories with symmetric tensors. Self-dualities with
symmetric tensors could be very interesting, since these theories
appear again in a different context: Pouliot's dualities \cite{pou} relate
$Spin(7,8,10)$ groups with spinorial matter and $SU$ groups with symmetric tensors.
It would be very interesting to see if this can be generalized to
higher $Spin$ groups, since there is no reason why 7, 8 and 10 should be special.
If this is true, the many self-dual $Spin$ groups we presented should
also have self-dual $SU$ groups as duals (if A is dual to B and
also self-dual, B has to be self-dual, too). Hence later ones would appear
as natural starting points for more dualities a la Pouliot,
which may lead to a deeper understanding of $N=1$ dualities.

In this light it is very disappointing that, as one can check easily,
there is no way to create a self-dual pair with the simple pattern
we described in this paper, at least as long as it only involves
one symmetric tensor and fundamentals and antifundamentals. Nevertheless
we can create new self-dualities with $SU$ groups, symmetric
tensors and an additional superpotential by using some of
our new self-dualities for groups that are known to also have 
a dual description in terms of a $SU$ theory and applying
the A dual B argument in the other direction.
We will demonstrate this at the example of $G_2$ with 6 flavors.
Besides the self-duality we presented, this is also dual to (\cite{pou})
\begin{center}
\begin{tabular}{|c||c||c|c|}
\hline
& $SU(3)$ & $SU(6)$ & $U(1)_R $\\
\hline
\hline
$q$ & $\overline{3}$ &
$\overline{ 6}$ & $\frac{1}{3}$  \\
\hline
$q^0$ & $\overline{3}$ & $1$ & $\frac{2}{3}$  \\
\hline
$s$ & ${6}$ & 1 & $\frac{2}{3}$ \\
\hline
\hline
$M$ & $1$ & $21$ & $\frac{2}{3}$ \\
\hline
\end{tabular}

with $W=q^0 q^0 s+ s^{N_f-3} + M qq s$.
\end{center}

Now this has to be self-dual, too. In fact it is very easy to see that
this is actually true, the dual being
\begin{center}
\begin{tabular}{|c||c||c|c|}
\hline
& $SU(3)$ & $SU(6)$ & $U(1)_R $\\
\hline
\hline
$Q$ & $\overline{3}$ &
$\overline{ 6}$ & $\frac{1}{3}$  \\
\hline
$Q^0$ & $\overline{3}$ & $1$ & $\frac{2}{3}$  \\
\hline
$S$ & ${6}$ & 1 & $\frac{2}{3}$ \\
\hline
\hline
$M$ & $1$ & $\tableau{2}$ & $\frac{2}{3}$ \\
\hline
$B$ & $1$ &$\tableau{1 1 1}$& $1$\\ 
\hline
\end{tabular}

with $W=Q^0 Q^0 S+ S^3 + M QQ S+ B Q^3$.
\end{center}

One can get rid of the field $M$ on both sides by giving it a mass. But
we have to live with the other superpotential terms. We created,
as advertised, a new self-dual pair of $SU$ theories with symmetric
tensor and tree level superpotential. A similar construction is possible
in other cases where we allready know a dual.

\section{Conclusion}

We found a criterion that allows us to find candidates for self-duality.
With this tool, we were able to create several new self-dual pairs,
including some with exceptional gauge group.
Even though for many of those theories this self-duality is all we
have so far, we strongly suspect that there often is yet another
dual description, also valid for more general matter content. Self-duality
might be used as a tool in constructing those dualities.


\begin{thebibliography}{99}

\bibitem{seiberg}
N. Seiberg, hep-th/9402044, {\it Phys. Rev} {\bf D49} (1995) 6857;
N. Seiberg, hep-th/9411149, {\it Nucl. Phys} {\bf B435} (1995) 129.
\bibitem{classical}
K. Intriligator and N. Seiberg, hep-th/9509066.
K. Intriligator and P.Pouliot, hep-th/9505006, {\it Phys. Let.}
{\bf 353B}, 471 (1995).
\bibitem{intrilligator}
K. Intriligator, R. Leigh and M. Strassler, hep-th/9506148.
\bibitem{kutasov}
D. Kutasov and A. Schwimmer, hep-th/9505004, {\it Phys. Lett.} {\bf B354}
(1995) 315.
\bibitem{spad}
R. Leigh and M. Strassler, hep-th/9505088, {\it Phys. Lett.} {\bf B356}
(1995) 492.
\bibitem{berkooz}
M. Berkooz, hep-th/9505067.
\bibitem{brodie}
J. Brodie and M. Strassler, hep-th/9611197;
J. Brodie, hep-th/9605232, {\it Nucl. Phys.} {\bf B478} (1996).
\bibitem{ramond}
P. Ramond, hep-th/9608077.
\bibitem{pou}
P. Pouliot, hep-th/9510148;
P. Pouliot, hep-th/9507018, {\it Phys. Lett.} {\bf B359} (1995) 108;
P. Pouliot and M. Strassler, hep-th/9510228;
P. Pouliot and M. Strassler, hep-th/9602031.
\bibitem{acci}
R. Leigh and M. Strassler, hep-th/9611020.
\bibitem{leigh}
R. Leigh and M. Strassler, hep-th/9503121, {\it Nucl. Phys.} {\bf B447}
(1995) 95.
\bibitem{luty}
M. Luty, M. Schmaltz, J. Terning, hep-th/9603034.
\bibitem{sp}
C. Cs\'aki, W. Skiba and M. Schmaltz, hep-th/9607210.
\bibitem{myfirst}
J. Distler and A. Karch, hep-th/9611088.
\bibitem{boston} C. Cs\'aki, M. Schmaltz, W. Skiba, J. Terning, hep-th/9701191.
\bibitem{confine} C. Cs\'aki, M. Schmaltz, W. Skiba, hep-th/9610139,
{\it Phys. Rev. Lett.} {\bf 78}, 799 (1997); hep-th/9612207.

\end{thebibliography}
\end{document}